\documentclass[a4paper,12pt]{article}
\usepackage{epsfig}
\begin{document}

\title{ Presence of Extra $P_{11}$ Resonances in Zagreb Analysis Since 1995.}

\author{S. CECI, A. \v{S}VARC and B. ZAUNER\\
Rudjer Bo\v{s}kovi\'{c} Institute, \\
Bijeni\v{c}ka c. 54, \\ 
10 000 Zagreb, Croatia\\ 
E-mail: Alfred.Svarc@irb.hr}

\maketitle

\begin{abstract}
The  partial wave T-matrices for the $\pi$N, $\eta$N and $\pi^2$N channels have been obtained within the framework of 
the coupled channel model using the $\pi$N elastic and $\pi$N $\rightarrow \eta$N data base as input. It has been 
shown that for the $P_{11}$ partial wave an equally good representation of the experimental data (namely the $T_{\pi 
N,\pi N}$ and $T_{\pi N,\eta N}$ T-matrices) can be obtained using either three, or four poles for the Green function 
propagator. However, the three Green function pole solution is not acceptable due to the structure of the extracted 
resonances. The two out of four $P_{11}$ resonances, those lying in the energy range 1700 MeV $< M_R <$ 1800 MeV, are 
poorly determined, but they seem to be strongly inelastic. The inclusion of other inelastic channels is needed to 
determine masses and widths of missing resonances with greater precision.
\end{abstract}

The  partial wave T-matrices for the $\pi$N, $\eta$N and $\pi^2$N channels have been obtained within the framework of 
the three body coupled channel model (CMU-LBL) using the $\pi$N elastic and $\pi$N $\rightarrow \eta$N data base as 
input\cite{Cut79,Bat98}. As it has been shown in Fig.1 the number of Green function poles (N), the respective pole 
positions ($s_i$) and the channel-resonance mixing parameters ($\gamma _{ai}$) are the input parameters of the 
fitting procedure which are adjusted in such a way that the experimental input ($\pi$N elastic T-matrices and $\pi$N 
$\rightarrow \eta$N data base) is well reproduced. In reality, the number of T-matrix poles (the Green function poles 
$s_i$), which are to be interpreted as resonant states,  is chosen in advance, and the only criterion is the quality 
of data reproduction. The schematic representation of the  fitting procedure is shown with full lines in Fig.1. Only 
upon obtaining the full coupled channel T-matrices set which satisfactory reproduces the experimental data,  the 
resonance parameters are extracted from the final set of obtained T-matrices.  Let us emphasize that the resonance 
parameters are  extracted {\em a posteriori}. We have no influence upon the type of resonance which we are going to 
obtain {\em during} the fitting procedure, we just know their number, namely, the number of Green function poles.

\begin{figure}[!h]
\begin{center} 
\epsfig{figure=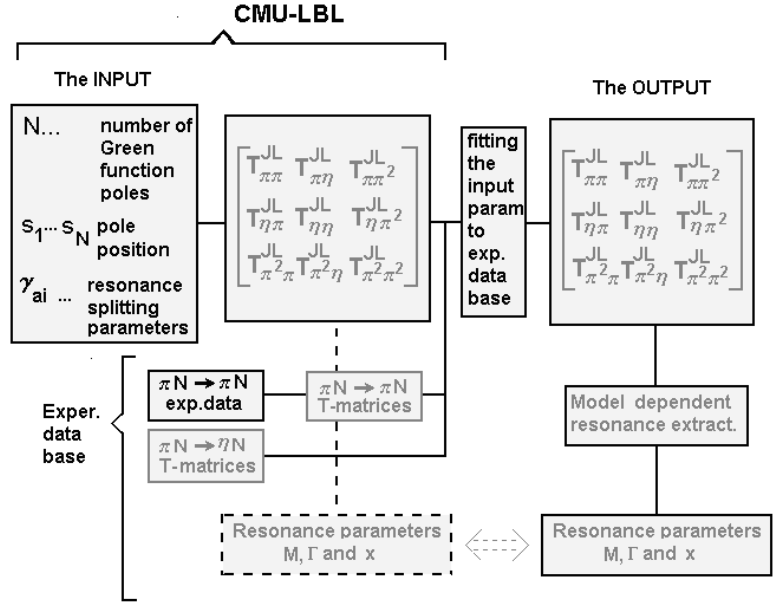,height=8cm}   
\caption{The schematic representation of resonance extraction in the coupled channel formalism (CMU-LBL). The full 
lines represent the present situation, the dashed lines are the suggested and needed modifications.}
\end{center}
\end{figure}

\begin{figure}[!h]
\begin{center}
\epsfig{figure=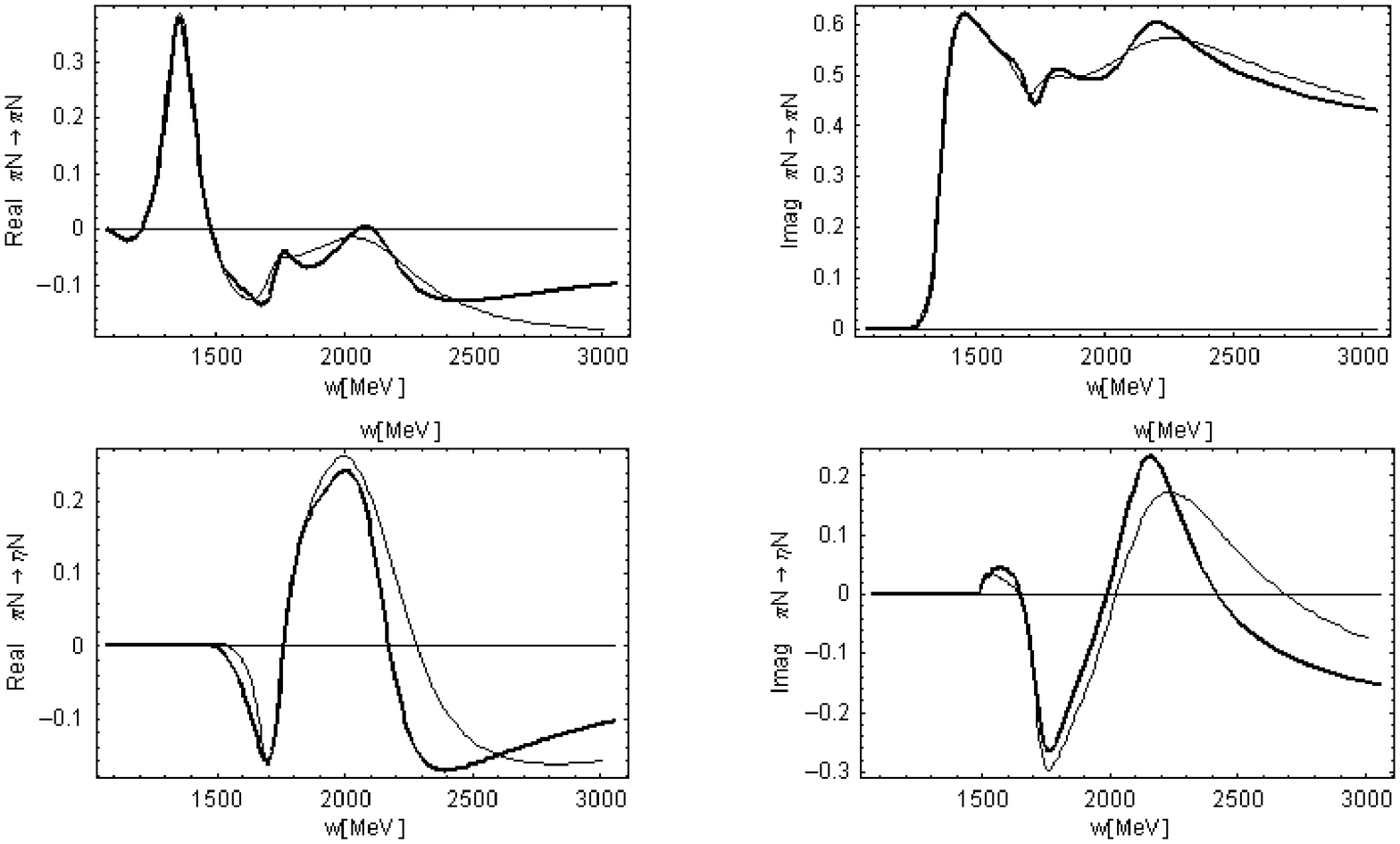,height=6cm}   
\caption{The $P_{11}$ partial wave T-matrices for three (thin lines) and four (thick lines) T-matrix poles for $\pi$N 
$\rightarrow \pi$N and $\pi$N $\rightarrow \eta$N processes.}
\end{center}
\end{figure} 

As it is shown in Fig.2. the same quality of the fit to the input experimental data base, which is reflected through 
almost identical form of the $T_{\pi N,\pi N}$ and $T_{\pi N,\eta N}$, is obtained with three (N=3, thin solid line) 
and four (N=4, thick solid line) poles in the Green function, respectively. However, as it is shown in Fig.3 the 
prediction for the remaining coupled channel T-matrices  $T_{\eta N,\eta N}$ and $T_{\pi^2 N,\pi^2 N}$ is 
dramatically different for the three and four pole solution (thick and thin line). The resonance parameters, 
extracted from both solutions are given in Table 1.

\begin{figure}[!h]
\begin{center}
\epsfig{figure=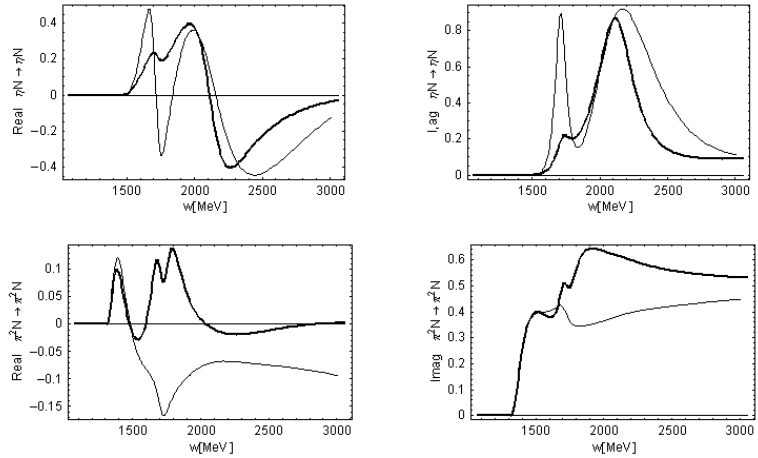,height=6cm}   
\caption{The $P_{11}$ partial wave T-matrices for three (thin lines) and four (thick lines) T-matrix poles for 
$\eta$N $\rightarrow \eta$N and $\pi^2$N $\rightarrow \pi^2$N processes.}
\end{center}
\end{figure}

We claim that the obtained three body solution is not acceptable because it shows strong coupling of second and third 
resonance to the $\eta$N channel {\bf only}, and shows {\bf no} branching ration to the third, effective channel 
which includes processes like $\pi$N $\rightarrow \pi \pi$N and $\pi$N $\rightarrow$ K $\Lambda$ which are 
experimentally 
firmly established.

\begin{table}[!h]
\begin{center}
Resonance parameters for the three and four pole solution.

{\footnotesize
\begin{tabular}{@{}||c|c|c|c|c|c||@{}}                \hline\hline
    States       & \multicolumn{5}{c||}{\bf Three poles in the Green function} \\
	 \cline{2-6} 
	   L$_{2I,2J}$                    &  Mass    &   Width    & $x_\pi$  & $x_\eta$ &  $x_{\pi^{2}}$ \\
  ${\rm (_{Mass/Width}^{x_{el}})}$ &  (MeV)   &   (MeV)    & (\%)     &   (\%)   &    (\%)       \\                                                                                                                                                   
\hline \hline
    P$_{11}(_{1440/135}^{51})$     & 1426(25)   &   287(53)    &  61(9)    &  0(0)        &      39(9)     \\
    P$_{11}(_{1710/120}^{12})$     & 1724(35)   &   116(47)    &  5(5)     &  {\bf 89(7)} &      6(5)  \\
    P$_{11} $                      & -          &   -          &   -       &  -           &      -     \\
    P$_{11}(_{2100/200}^{9})$       & 2175(89)  &   659(207)   &   9(4)   &  {\bf 89(3)}  &     2(2)  \\ \hline 
\hline
States &   \multicolumn{5}{c||}{\bf Four poles in the Green function} \\
             \cline{2-6}
			 L$_{2I,2J}$                    &  Mass    &   Width    & $x_\pi$  & $x_\eta$ &  $x_{\pi^{2}}$ \\ 
			 ${\rm (_{Mass/Width}^{x_{el}})}$   &  (MeV)   &   (MeV)    & (\%)     &   (\%)   &    (\%)       \\
\hline \hline
   P$_{11}(_{1440/135}^{51})$   &   1439(19)   &    437(141)     &   62(4)     &     0(0)      &    38(4)         \\
   P$_{11}(_{1710/120}^{12})$   &   1729(16)   &    180(17)      &    22(24)   &     6(8)      &   {\bf 72(23)}     
\\
   P$_{11} $                    &   1740(11)   &    140(25)      & 28(34)      &    12(9)      &    {\bf 60(35)}    
\\ 
   P$_{11}(_{2100/200}^{9})$    &   2157(42)  &    355(88)    & 16(5)      &  {\bf 83(5)}          &    1(1)     \\ 
\hline \hline
\end{tabular} }
\end{center}
\end{table} 
 We offer two alternative explanations: either our fitting procedure is technically inadequate to find a better three 
pole solution, or we indeed need four resonances in the $P_{11}$ partial wave, second and third strongly inelastic, 
exactly as indicated in Table 1.

We conclude that using $\pi$N elastic and $\pi$N $\rightarrow \eta$N data base is, at the present moment, 
insufficient even to determine the $\eta$N elastic channel. 

In order to improve the fitting technique, we propose to include the resonance parameters as the "quasi input" into 
the fitting procedure (dashed lines in Fig.1). That would enable us to search for a particular type of a three body 
solution, namely solution which reproduces the experimental data set {\bf simultaneously} with imposing that one of 
the  resonances is inelastic in other then $\eta$N channel. If such a procedure fails in the end, the statement that 
we need more then three resonances in a $P_{11}$ partial wave is fully justified.

\end{document}